\documentclass[aps,twocolumn,pra,superscriptaddress,amsmath,showpacs,tightenlines]{revtex4-1}
\usepackage{amssymb}
\usepackage{amsmath}
\usepackage{graphicx}
\usepackage{subfigure}
\usepackage{natbib}
\usepackage{epsfig}
\usepackage{amsfonts}
\usepackage{mathrsfs}
\usepackage{CJK}
\usepackage{xcolor}
\usepackage[toc,page,title,titletoc,header]{appendix}
\begin{document}

\title{Entanglement preparation and non-reciprocal excitation evolution in giant atoms by controllable dissipation and coupling}
\author{Hongwei Yu}
\affiliation{Center for Quantum Sciences and School of Physics, Northeast Normal University, Changchun 130024, China}
\author{Zhihai Wang}
\email{wangzh761@nenu.edu.cn}
\affiliation{Center for Quantum Sciences and School of Physics, Northeast Normal University, Changchun 130024, China}
\author{Jin-Hui Wu}
\affiliation{Center for Quantum Sciences and School of Physics, Northeast Normal University, Changchun 130024, China}

\begin{abstract}
We investigate the dynamics of giant atom(s) in a waveguide QED scenario, where the atom couples to the coupled resonator waveguide via two sites. For a single giant atom setup, we find that the atomic dissipation rate can be adjusted by tuning its size. For the two giant atoms system, the waveguide will induce the controllable individual and collective dissipation as well as effective inter-atom coupling. As a result, we can theoretically  realize the robust entangled state preparation and non-reciprocal excitation evolution. We hope our study can be applied in quantum information processing based on photonic and acoustic waveguide setup.
\end{abstract}

%\date{\today}

\maketitle

\section{Introduction}
The giant atom, which can be realized by superconducting qubits (artificial atoms), is a new component in the field of quantum optics~\cite{TP2003}. Since the size of the giant atom can be comparable to the wavelength of light, the traditional dipole approximation breaks down in light-atom interaction. {As a result,} the giant atom is nonlocally coupled to the waveguide via multiple connecting points and the interference effect between these points will dramatically modulate the collective behavior of the atoms. In this community, the mutual control between photon and atom is now attracting more and more attentions, both theoretically and experimentally~\cite{AF2014,LG2017,AF2018,PT,GA2019,AG2019,LG2019,SG2019,BK2019,AM2021}.

The coupled resonator waveguide (CRW) is widely studied in photon based quantum network~\cite{HJ2008}. On the one hand, it supplies a channel for the travelling photon with tunable group velocity, and the single~\cite{Fan2005,DE2007} or few photon scattering~\cite{PL2010,TS2018} has been used to construct the photon device,  such as quantum transistor~\cite{Zhou2008}, {router}~\cite{Zhou2013} and frequency converter~\cite{Wang2014}. On the other hand, due to its exotic energy band, the CRW provides a structured environment for the atom, to form the atom-photon dressed state and control the dissipation or decoherence of the atom~\cite{GC2016}. Meanwhile, as a data bus, the CRW can also induce the effective coupling between remote atoms~\cite{ES2013,PF2016,HZ2013}, and is therefore widely used in quantum information processing.

In the giant atom-CRW coupled system, we have shown that the size of the giant atom can serve as a sensitive controller, to regulate the single-photon transmission and photonic bound state in the waveguide~\cite{Zhao2020}. The further
question is how to control the dissipation and indirect interaction between the giant atom(s) by its (their) own size when {being subject to the environment which is}  composed by the CRW.

To tackle this issue, we first consider a system composed by a  {single} giant atom, which couples to a CRW via two distant sites. Within the Born-Markovian approximation, we show that the dissipation rate of the giant atom can be equal {to} or smaller than that for conventional small atom, or even surprisingly  achieves zero, depending on the size of the atom. It is then generalized to the system composed by two giant atoms, we find that the collective decoherence and effective interaction {are} accompanied by the individual dissipation, and all of these processes are finely controllable. By making all of the waveguide induced dissipation to be zero, we propose a robust scheme to realize the entanglement preparation. More interestingly, when only the collective dissipation is suppressed, we can realize a parity-time($\mathcal{PT}$)-like symmetry and non-reciprocal excitation {evolution for a fixed giant atom}. However, the excitation transmission between them is on contrary reciprocal, which is dramatically different from that in the quantum system with time reversal symmetry broken~\cite{atomic1,atomic2, PT1,PT2}.

The rest of the paper is organized as follows. In Sec.~\ref{single}, we present the single giant atom model and discuss its controllable dissipation due to the coupling to the waveguide. In Sec.~\ref{two}, we {generalize} to the two atom setup and investigate the applications in entangled state preparation and non-reciprocal excitation {evolution}. In Sec.~\ref{con}, we give a short conclusion and some remarks.

\section{Controllable dissipation for single giant atom}
\label{single}

\begin{figure}
  \centering
  \includegraphics[width=1\columnwidth]{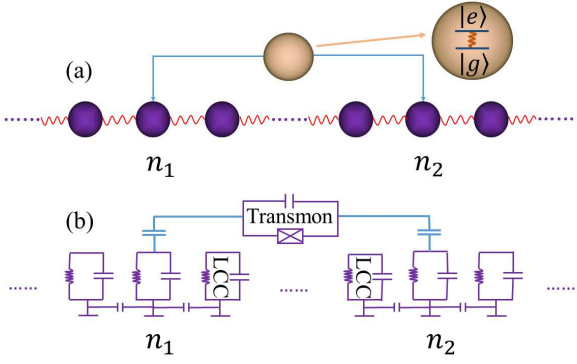}
  \caption{(a) Sketch of the waveguide QED setup, where a giant atom
is coupled to a photonic lattice via $n_1$th and $n_2$th sites. (b) The effective circuit diagram of the device.}
\label{figurepet}
\end{figure}

{As sketched in Fig.~\ref{figurepet} (a),  the system we consider is composed by an array of $N_c\rightarrow\infty$ coupled-resonator waveguide and a two-level system.} Here, we consider a giant atom scenario, where the two-level system couples to the waveguide via two sites. In what follows of this work, we will name such a two-level system as ``giant atom''. {The considered atom-waveguide coupled system can be realized in the superconducting quantum circuits which is demonstrated in Fig.~\ref{figurepet} (b). Here, the LC circuits (LCC) serve as the resonators, and the transmon qubit serves as the two-level system, that is, the giant atom. The capacities couple the resonators as well as the resonators and the transmon.}
{The Hamiltonian of the system is written as $H=H_c+H_I$, where ($\hbar=1$)
\begin{eqnarray}
H_{c}&=&\omega_{c}\sum_{j}a_{j}^{\dagger}a_{j}-\xi\sum_{j}\left(a_{j+1}^{\dagger}a_{j}+a_{j}^{\dagger}a_{j+1}\right),\\
H_I&=&\Omega|e\rangle \langle e|+g[(a_{n_{1}}^{\dagger}+a_{n_{2}}^{\dagger})\sigma^{-}+{\rm H.c.}].
\end{eqnarray}}
{Here,} $\omega_c$ is the frequency of the resonators, and $a_j$ is the bosonic annihilation operator on site $j$. $\xi$ is the hopping strength between the nearest resonators. $\sigma^{\pm}$ are the usual Pauli operators of the giant atom, and $\Omega$ is the transition frequency between the ground state $|g\rangle $ and the excited state $|e\rangle$.  {Here, we have considered that the giant atom couples to the waveguide via $n_1$th and $n_2$th resonators with coupling strength $g$.}  We have performed the rotating wave approximation in the inter-resonator and atom-waveguide coupling Hamiltonian.

Introducing the Fourier transformation $a_j=\sum_k a_k e^{ikj}/\sqrt{N_c}$, the Hamiltonian of the waveguide $H_c$ can be written in a diagonal form $H_c=\sum_k\omega_k a_k^\dagger a_k$ where the dispersion relation is given by $\omega_k=\omega_c-2\xi\cos k$. The waveguide therefore supports a single photon continual band which centers at $\omega_c$ with a width of $4\xi$. In this sense, the waveguide supplies a structured environment for the giant atom. When the giant atom is outside the waveguide in frequency domain, the dissipation will be suppressed due to the dispersive coupling to the waveguide. On the contrary, when the transition frequency of the giant atom locates inside the waveguide and far away from the upper and lower edges of the band, the initial excited atom will undergo an exponential decay in population. {In this sense, the frequency of the giant atom can be used to control its dissipation dynamics via the trivial resonant mechanism. However, we will show here that the interference effect induced by the intrinsic character of the giant atom makes its size another controller, even when the atomic frequency is inside the energy band of the waveguide, for example, the giant atom is resonant with the bare resonator in the waveguide.}

To obtain the master equation for the density matrix of the giant atom, we work in the momentum representation. Then, the atom-waveguide coupling Hamiltonian is expressed as~\cite{GC2016}
 \begin{equation}
H_I(t)=g\sum_{i=1}^{2}[\sigma^{+}E(n_{i},t)e^{i\Omega t}+\sigma^{-}E^{\dagger}(n_{i},t)e^{-i\Omega t}]
\label{interaction}
\end{equation}
where $E(n_{i},t)=\frac{1}{\sqrt{N_{c}}}\sum_{k}(e^{-i\omega_{k}t}e^{ikn_{i}}a_{k})$. With the Born-Markovian approximation, the master equation is formally written as~\cite{book}
\begin{equation}
\dot{\rho}(t)=-\int_{0}^{\infty}d\tau {\rm Tr}_{c}\{[H_I(t),[H_{I}(t-\tau),\rho_{c}\otimes\rho(t)]]\}.
\end{equation}

In what follows of this paper, we will consider that the giant atom is resonant with the bare resonator, that is $\Omega=\omega_c$. In this situation, after some direct calculations as shown in Appendix~\ref{master}, the master equation is finally simplified as
\begin{equation}
\dot{\rho}=-i\Omega[|e\rangle\langle e|,\rho]+(A+A^{*})\sigma^{-}\rho\sigma^{+}-A\sigma^{+}\sigma^{-}\rho-A^{*}\rho\sigma^{+}\sigma^{-}
\label{masters}
\end{equation}
where
\begin{equation}
A=\frac{g^{2}}{\xi}\left(1+e^{i\pi N/2}\right),
\label{CA}
\end{equation}
with $N=|n_1-n_2|$ characterizing the size of the giant atom, {that is, the distance between the two atom-CRW connecting points.}

It shows that the dissipation of the giant atom can be tuned on demand by changing its size. For example, when $N=4m+2$ with integral $m=0,1,2\cdots$, the dissipation rate is $A=0$, which implies that the atom will not undergo dissipation and decoherence.  Therefore, we can realize a decoherence protection via a giant atom setup even when it locates inside the waveguide in energy and this protection is not possible in a traditional small atom scheme, which interacts with only one site. On the other hand, when $N=4(m+1)$, the dissipation rate of the giant atom becomes {$J=2g^2/\xi$}, which is same to that in small atom setup when the atom-waveguide coupling strength is $2g$. At last, when the size of the giant atom satisfies $N=2m+1$, {$A=g^2(1\pm i)/\xi$} becomes a complex number, with the real part representing the decay rate {$J_0=J/2$} and the imaginary part {$\delta_0=\pm J_0$} representing the atom-waveguide coupling {inducing} frequency shift. {In the typical waveguide system composed by superconducting circuits, where the inter-resonator coupling strength can achieve by $\xi/(2\pi)=100$~MHz~\cite{SH2015}, it is easy to work in the parameter regime of $\delta_0\ll|2\xi|$, where the modified frequency of the giant atom is still inside the single photon band of the waveguide, and we can still describe the dynamics of the atom via the master equation in Eq.~(\ref{masters}).}

\begin{figure}
  \centering
\includegraphics[width=1\columnwidth]{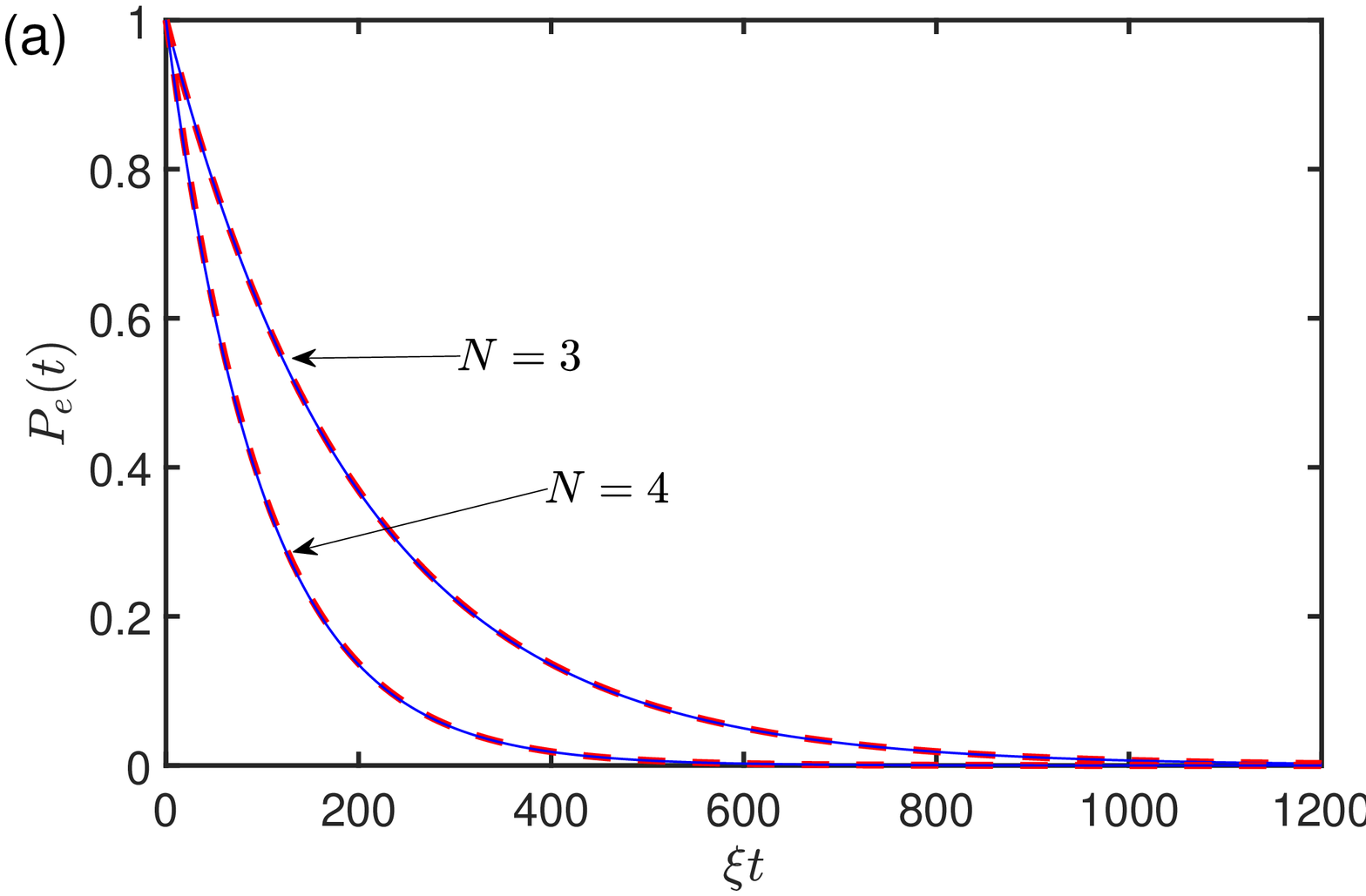}
\includegraphics[width=1\columnwidth]{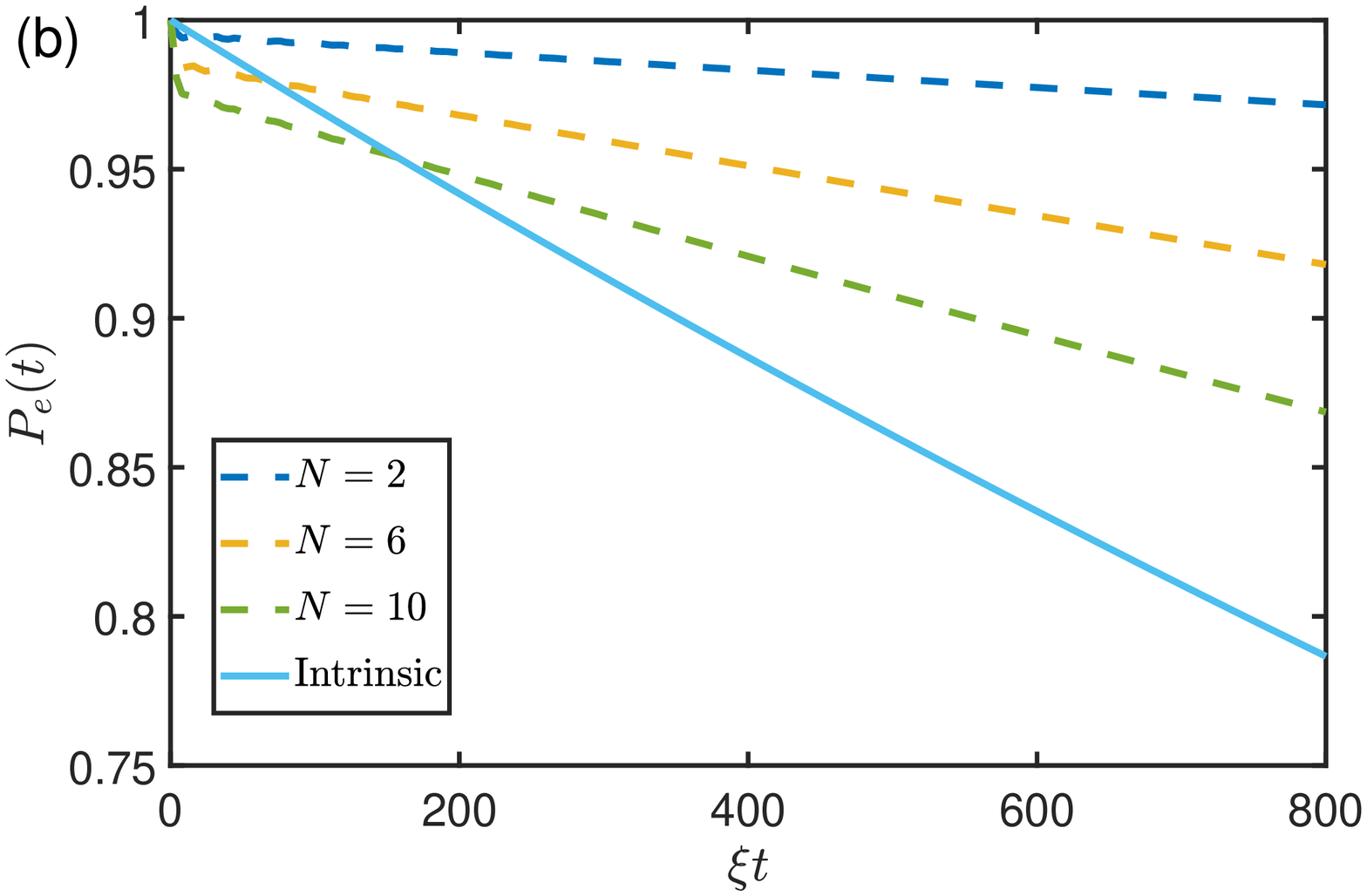}
\includegraphics[width=1\columnwidth]{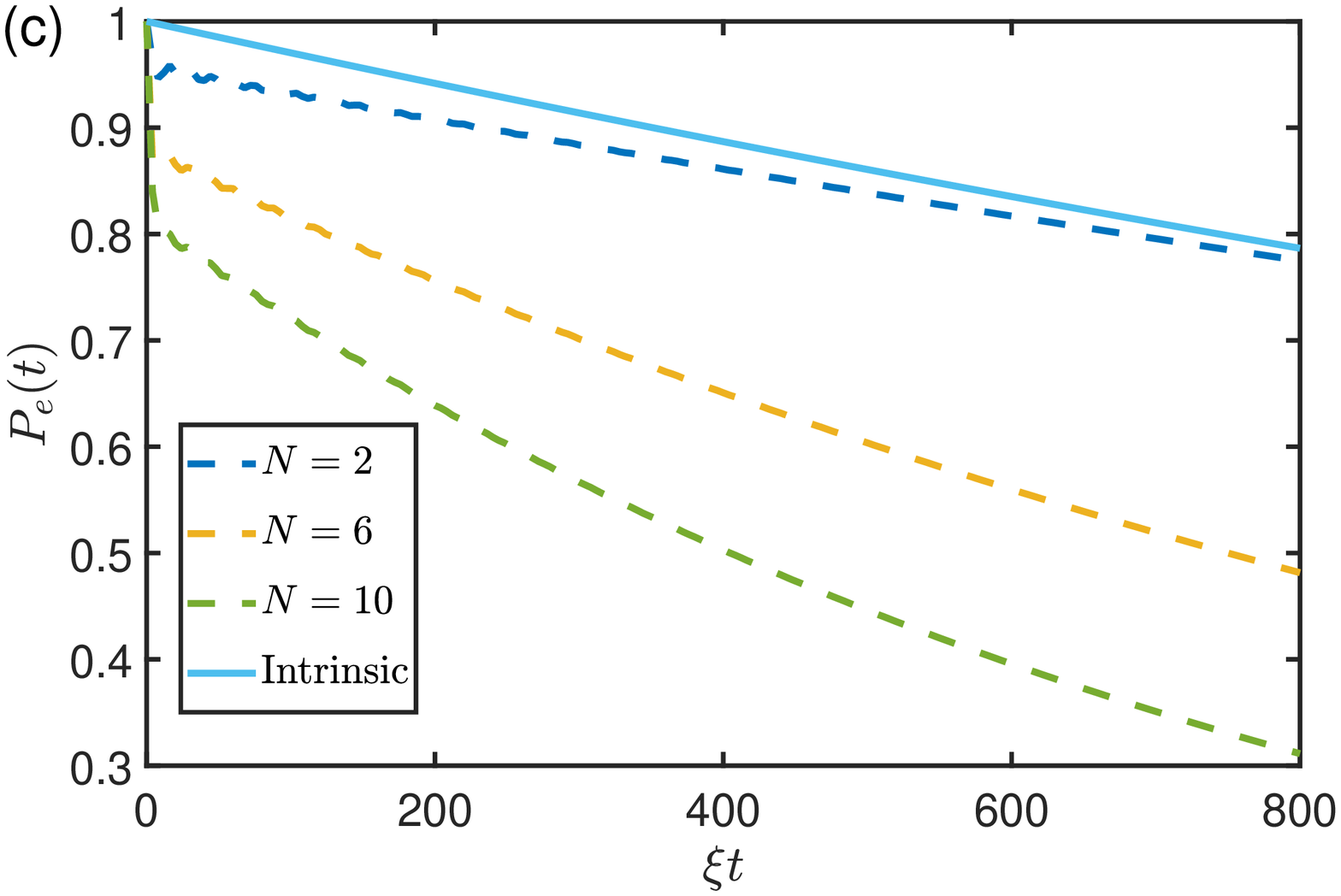}
  \caption{{Evolution of the excited state population $P_e(t)$. The dashed and solid lines in (a) are respectively the approximate analytic results in Eq.~(\ref{pet}) and the exact numerical results (see Appendix B).  The parameters are set as $\Omega=\omega_c$, $N_c=4000, \kappa/\xi=6\times10^{-3}$.  {$g/\xi=0.05$ for (a,b) and $g/\xi=0.15$} for (c). The intrinsic decay rate for the giant atom is taken as $\gamma_{1}/\xi=3\times10^{-4}$ in (b) and (c).} }
\label{atomdecay}
\end{figure}

{The underlying physics in the above discussion is that the waveguide serves as a structured environment, which induces the dissipation of the giant atom.  As a result, the evolution of the excited state population yields
 \begin{equation}
 P_e(t)=|\langle |e\rangle\langle e|\rangle|=e^{-2{\rm Re}(A)t}
 \label{pet}
 \end{equation}
 for the initial excited giant atom.

For $N=3$ and $N=4$ where ${\rm Re} (A) \neq 0$, we plot the curve of $P_e(t)$ by neglecting the small atomic intrinsic dissipation in Fig.~\ref{atomdecay} (a). Here, the solid lines are the approximate results obtained by Eq.~(\ref{pet}) and the dashed lines are the numerical results by including a small loss rate $\kappa$ for each resonator {(see Appendix~\ref{numer} for the detailed description of the numerical calulations)}. The results in Fig.~\ref{atomdecay} (a) show  a good agreement between the approximate and the numerical treatment for the atom-waveguide coupling strength {$g=0.05\xi$}. Furthermore, Eq.~(\ref{pet}) implies that $P_e(t)=e^{-2g^2 t/\xi}$ for $N=3$ and $P_e(t)=e^{-4g^2 t/\xi}$ for $N=4$}. Therefore, the atom for $N=4$ decays faster than that for $N=3$, which is also clearly demonstrated in Fig.~\ref{atomdecay} (a). Our further simulations (not shown here) show that Eq.~(\ref{pet}) works well even when the atom-waveguide coupling strength achieves {$g/\xi=0.2$}, below which the numerical result is nearly independent of $N$.

{On the other hand, for $N=2,6,10...$, the Markovian master equation in Eqs.~(\ref{masters},\ref{CA}) tells us the atom will not decay by coupling to the waveguide channel. Thus, the main dissipation comes from intrinsic decay, which is induced by the coupling to the surrounding environment except the waveguide. Then, the population yields $P_e(t)=e^{-\gamma_1 t}$, where $\gamma_{1}$ is the intrinsic decay rate. It is therefore necessary to investigate when the high-order non-Markovian effects will dominate the intrinsic decay. To this end, we show the full numerical dynamics for the above $N$s in Figs.~\ref{atomdecay} (b) and (c) as well as the dynamics with only the intrinsic dissipation for comparisons. {Here, the dashed lines are the results by omitting the intrinsic decay, that is, the giant atom does not couple to the other environments except for the waveguide. The solid lines are the results by only considering the intrinsic decay, but without that induced by the waveguide channel.} It is obvious that the numerical dynamics (dashed lines) depends dramatically on $N$, and the excited state population decay becomes faster as $N$ increases. For a weak atom-waveguide coupling strength {($g/\xi=0.05$)} in Fig.~\ref{atomdecay} (b), the decay induced by the waveguide channel for $N=2,6,10$ is dominated by the intrinsic decay in the long time scale with $\gamma_1/\xi=3\times10^{-4}$, {that is, the dashed lines are above the solid lines}. By increasing $g$ while fixing $\gamma_1$, we show in Fig.~\ref{atomdecay} (c) that the waveguide channel induced decay will surpass intrinsic decay even for $N=2$, which is the smallest value allowed by $A=0$, for $g/\xi=0.15$. In this situation, the high-order or non-Markovian effect becomes more important than the error source intrinsic to the giant atom, and {the dashed lines are below the solid lines.}

\section{Two giant atom setup}
\label{two}

To explore the potential application of giant atom in the quantum information processing, we generalize the above discussions to the setup consisting {of} two giant atoms. Then, the total Hamiltonian for the system is
{\begin{eqnarray}
H_{2}&=&H_{c}+\Omega(|e\rangle_{1}\langle e|+|e\rangle_{2}\langle e|)\nonumber \\&&+g(a_{n_{1}}^{\dagger}\sigma_{1}^{-}+a_{n_{2}}^{\dagger}\sigma_{1}^{-}+{\rm H.c.})\nonumber
\\&&+g(a_{m_{1}}^{\dagger}\sigma_{2}^{-}+a_{m_{2}}^{\dagger}\sigma_{2}^{-}+{\rm H.c.}),
\end{eqnarray}}
which implies that the first atom couples to the waveguide at the $n_1$th and $n_2$th sites, while the second atom couples to the waveguide at the $m_1$th and $m_2$th sites. Similar to the treatment for the single giant atom, we can obtain the master equation as
\begin{eqnarray}
\dot{\rho}=-i[\mathcal{H},\rho]+\sum_{i,j=1}^{2}\frac{\Gamma_{ij}}{2}(2\sigma_{j}^{-}\rho\sigma_{i}^{+}-\sigma_{i}^{+}\sigma_{j}^{-}
\rho-\rho\sigma_{i}^{+}\sigma_{j}^{-}), \nonumber \\
\end{eqnarray}
where the coherent coupling between the two atoms is described by the Hamiltonian
\begin{equation}
\mathcal{H}=\sum_{i=1}^2(\Omega+\frac{U_{ii}}{2})|e\rangle_i\langle e|+\frac{U_{12}}{2}(\sigma_1^+\sigma_2^-+{\rm H.c.}).
\end{equation}
In the above equations, we have defined $U_{ij}:=2{\rm Im}(A_{ij}), \Gamma_{ij}:=2{\rm Re}(A_{ij})$ and
{\begin{subequations}
\begin{eqnarray}
A_{11}&=&\frac{g^{2}}{\xi}(1+e^{i\frac{\pi}{2}\left|n_{1}-n_{2}\right|}),\\
A_{22}&=&\frac{g^{2}}{\xi}(1+e^{i\frac{\pi}{2}\left|m_{1}-m_{2}\right|}),\\
A_{12}&=&A_{21}=\frac{g^{2}}{2\xi}\sum_{i,j=1}^{2}e^{i\frac{\pi}{2}\left|n_{i}-m_{j}\right|}.
\end{eqnarray}
\label{CAS}
\end{subequations}}

It shows that the individual dissipation rate ($\Gamma_{11}$ and $\Gamma_{22}$ ) and frequency shift ($U_{11}$ and $U_{22}$ ) of the two giant atoms are determined by the size of each giant atom.  Furthermore, the waveguide can also serve as a data bus, to induce the interaction and collective dissipation between two giant atoms and the rates are given by $U_{12}$ and $\Gamma_{12}$, respectively. They are determined by both of the size of the two giant atoms and their relative locations. Taking the advantage of tunable nature of $A_{ij}$ by the formulation of the giant atom, we will show two applications in quantum information processing in what follows.

\subsection{Entangled state preparation}
As the first application, we discuss the entangled state preparation. We can appropriately choose $n_1, n_2, m_1, m_2$ such that it satisfies $\Gamma_{11}=\Gamma_{22}=U_{11}=U_{22}=\Gamma_{12}=0$ and {$U_{12}=2J=4g^2/\xi$}. In this situation, the waveguide will only induce the interaction between the two giant atoms. As a result, the master equation in the rotating frame defined by the free term of atoms is reduced into $\dot{\rho}=-i[H_0,\rho]$
with
{\begin{equation}
H_0=J(\sigma_{1}^{+}\sigma_{2}^{-}+\sigma_{2}^{+}\sigma_{1}^{-}).
\label{HH}
\end{equation}}
Preparing the initial state as $|\psi(0)\rangle=\left|e;g\right\rangle$, which represents that the first atom is in the excited state while the second one is in the ground state. At an arbitrary moment $t$, the wave function of two giant atom system becomes
{\begin{equation}
 \left|\psi\right\rangle=\cos(J t)\left|e;g\right\rangle-i\sin(J t)\left|g;e\right\rangle.
\end{equation}}
Choosing the evolution time {$t=\pi/(4J)$}, we can achieve the maximum entangled state {$\left|\psi\right\rangle=(\left|e;g\right\rangle-i\left|g;e\right\rangle)/\sqrt{2}$}.

In the above entangled state preparation scheme, we have only considered the effect of the waveguide. In fact, the atoms also inevitably interact {with} the external environment. In such case, the dynamics of the system is governed by the master equation
\begin{eqnarray}
\dot{\rho}&=&-i[H_{0},\rho]\nonumber \\&&+\frac{\gamma_{1}}{2}\sum_{i=1}^{2}(2\sigma_{i}^{-}\rho\sigma_{i}^{+}-\sigma_{i}^{+}\sigma_{i}^{-}\rho
-\rho\sigma_{i}^{+}\sigma_{i}^{-})\nonumber \\&&+\frac{\gamma_{2}}{2}(\sigma_{1}^{z}\rho\sigma_{1}^{z}
+\sigma_{2}^{z}\rho\sigma_{2}^{z}-2\rho).
\end{eqnarray}
Here, the first line represents the unitary evolution mediated by the waveguide. The second line represents the dissipation/decoherence process with rate $\gamma_1$, which is same with that in the last section and the third line represents the pure dephasing process with rate $\gamma_2$.

In the superconducting circuits, the LCCs serve as the resonators and the transmon serves as the giant atom. In the recent experiments, the coupling strength between the nearest LCCs can be achieved by $\xi/(2 \pi)=100$~MHz~\cite{SH2015}, and the light-matter interaction have achieved the ultra- and deep-strong coupling regime in superconcuting circuit QED system~\cite{PF2017,FY2017}. The transition frequency of the giant atom and the eigen frequency of the LCC are both in the order of {several GHZ, and} the atom-waveguide coupling strength is easily to {achieve the} order of MHz, so the waveguide {inducing} atom-atom coupling strength $J=2g^2/\xi$ is in the order of MHz. Moreover, the intrinsic decoherence time of the superconducting qubit (for example the transmon) is achieved by $T_1=20~{\rm \mu s}$ and $T_2^{*}=10~{\rm \mu s}$~\cite{AJ2017} or even longer~\cite{JJ2019}, which implies $\gamma_1/(2\pi)\leq 8$~kHz and $\gamma_2/(2\pi)\leq16$~kHz, that is, $\gamma_1/J$ and $\gamma_2/J$ are in the order of $10^{-3}\sim10^{-2}$. In Fig.~\ref{entanglementf} (a), we plot the fidelity as a function of $\gamma_1$ and $\gamma_2$ based on the above master equation. It shows that, even for $\gamma_1=0.1J, \gamma_2=0.2J$, the fidelity is still higher than $92\%$.

\begin{figure}
  \centering
\includegraphics[width=1\columnwidth]{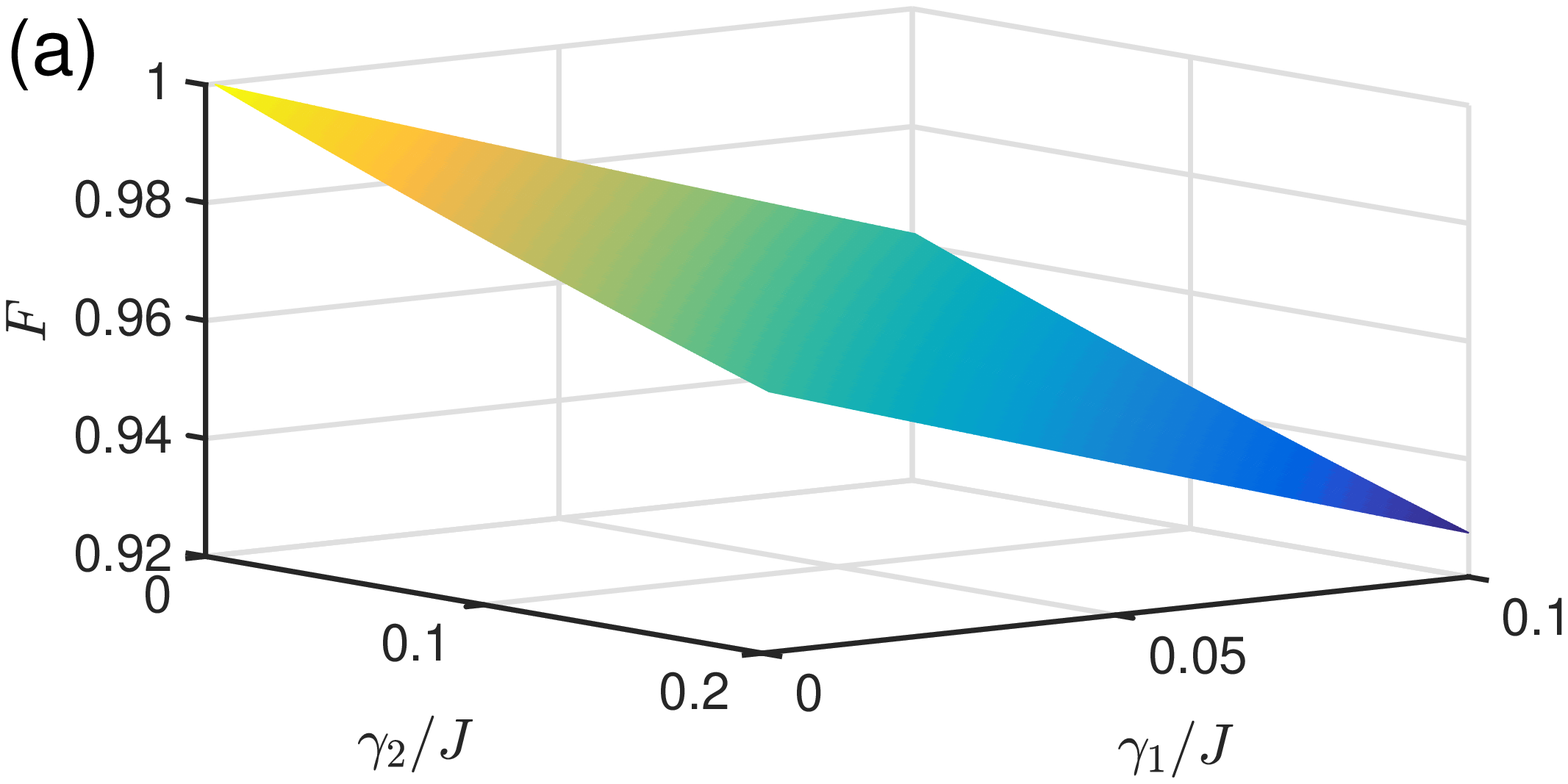}
\includegraphics[width=1\columnwidth]{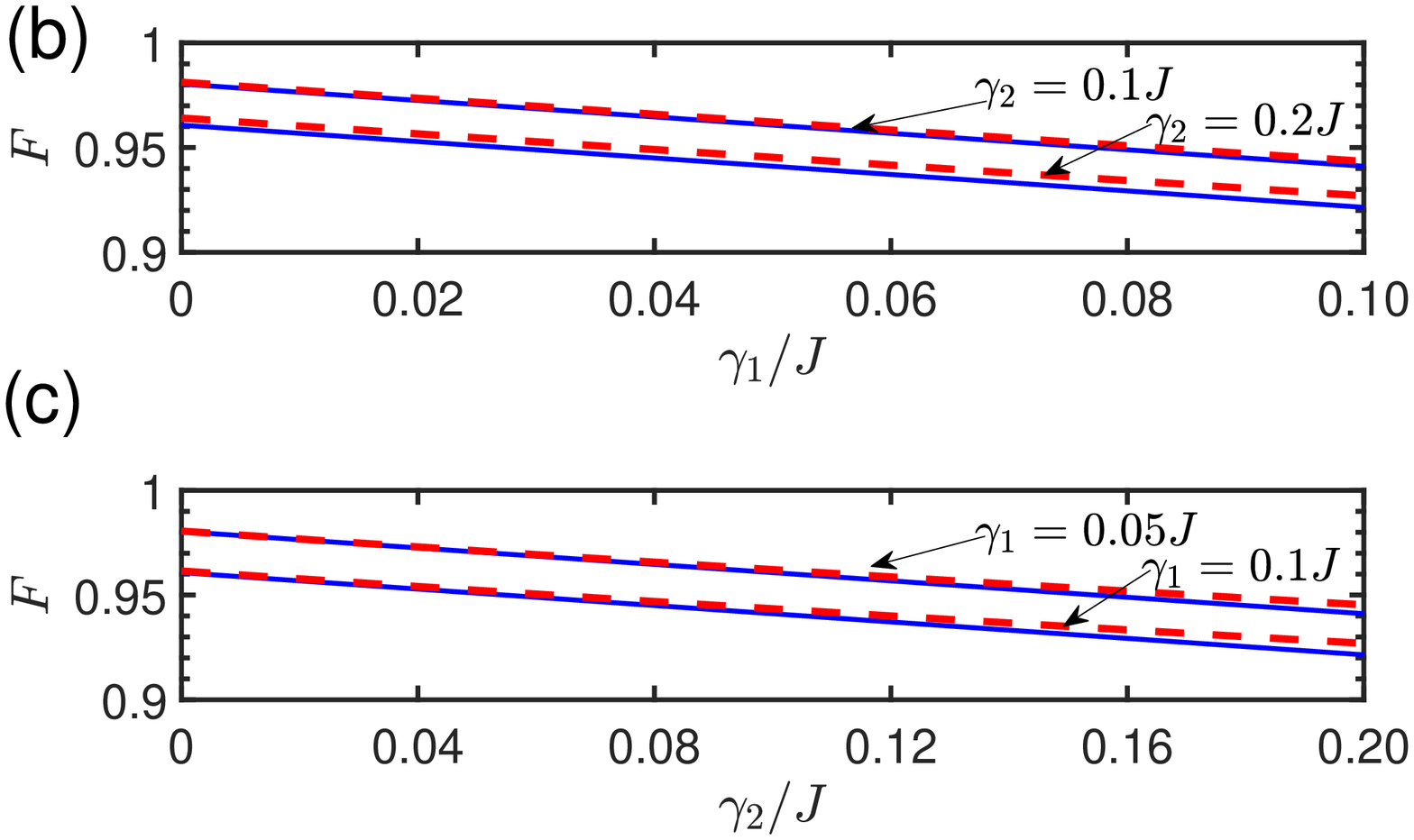}
  \caption{ The fidelity of entangled state preparation. (a) The fidelity based on the master equation. (b,c) The fidelity as functions of $\gamma_1$ ($\gamma_2$) for different $\gamma_2$ ($\gamma_1$). Here, $\gamma_1$ and $\gamma_2$ are given in the unit of {$J$}.}
\label{entanglementf}
\end{figure}

In our consideration, the effect of the external environment is much smaller than that induced by the waveguide, {that is, $(\gamma_1,\gamma_2)\ll J$. Up to the first order of $\gamma_1/J$ and $\gamma_2/J$, the fidelity for the entangled state preparation is approximated as
\begin{equation}
F=\sqrt{\left\langle\psi\right|\rho(t=\frac{\pi}{4J})\left|\psi\right\rangle}\approx 1-\frac{\pi}{8J}\gamma_{1}-\frac{\pi}{16J}\gamma_{2}.
\label{fide}
\end{equation}}
It shows that the coupling to the external environment is harmful for the entangled state preparation in our system and the decoherence process dominates the pure dephasing process in decreasing the fidelity. In Figs.~\ref{entanglementf} (b,c), we compare the fidelity obtained from the master equation (solid lines) and that from the approximated expression (dashed lines) in Eq.~(\ref{fide}). The fidelity shows a linear dependence on $\gamma_1$ as well as $\gamma_2$, and the agreement between them implies the validity of Eq.~(\ref{fide}).

{In the above discussion, we have chosen the fixed geometry configuration for the giant atoms, that is, the fixed size for each giant atom and distance between them. In such situation, the waveguide only induces the effective interaction between the two giant atoms, but without dissipations. One may also investigate the steady state entanglement by including a continuous weak driving laser, such that the entanglement will show an oscillatory dependence on the size of the giant atom and the distance between them. The underlying physics is similar to that of two small atoms setup~\cite{HZ2013}. However, such driven-dissipation induced entanglement is beyond the discussion in this work.}

\subsection{{Non-reciprocal excitation evolution}}

Next, let us discuss the application in demonstrating the $\mathcal{PT}$-like symmetry physics and non-reciprocal excitation {evolution}. On contrary to the entangled state preparation scheme, we here set
$U_{11}=U_{22}=\Gamma_{12}=0$ and {$U_{12}=2J$}, that is, the two atoms effectively interact with each other
and undergo individual dissipations. The dynamics is then governed by the master equation  (in the rotating frame)
\begin{eqnarray}
  \dot{\rho}&=&-i[H_{0},\rho]+\frac{\Gamma_{11}}{2}(2\sigma_{1}^{-}\rho\sigma_{1}^{+}
  -\sigma_{1}^{+}\sigma_{1}^{-}\rho-\rho\sigma_{1}^{+}\sigma_{1}^{-})\nonumber \\
&&+\frac{\Gamma_{22}}{2}(2\sigma_{2}^{-}\rho\sigma_{2}^{+}-\sigma_{2}^{+}\sigma_{2}^{-}\rho
-\rho\sigma_{2}^{+}\sigma_{2}^{-}),
\label{master2}
\end{eqnarray}
where the Hamiltonian $H_0$ is given in Eq.~(\ref{HH}).

We now consider that only one atom is excited initially, and {explore the evolution of the excited probability at the same atom.} When the initial excitation is at the first atom, that is, $|\psi_1(0)\rangle=|e;g\rangle$, the {excited probability} is obtained as

{\begin{eqnarray}
P_1(t)&=&{\rm Tr}[|e\rangle_1\langle e|\rho_1(t)]\nonumber \\
&=&\frac{e^{-\frac{\delta t}{2}}}{2K}
[(\Delta^2-8J^2)M_+(t)+\sqrt{K}\Delta M_-(t)-16J^2]\nonumber \\
\label{trap1}
\end{eqnarray}}
where $\rho_1(t)$ is the density matrix at time $t$, $\delta=\Gamma_{11}+\Gamma_{22}, \Delta=\Gamma_{11}-\Gamma_{22}$,
and

{\begin{eqnarray}
K&=&\Delta^2-16J^2,\\
M_{\pm}(t)&=&e^{-\frac{\sqrt{K}t}{2}}\pm e^{\frac{\sqrt{K}t}{2}}.
\end{eqnarray}}

\begin{figure}
  \centering
\includegraphics[width=1\columnwidth]{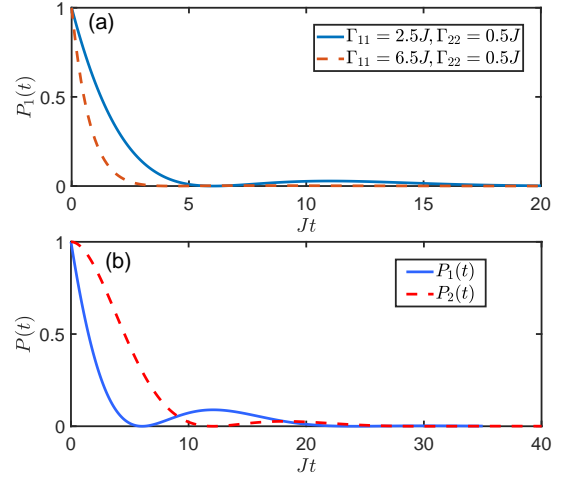}
\caption{(a) The trapping population $P_1(t)$. (b) The demonstration of non-reciprocal excitation trapping. The parameters for (b) is set as $\Gamma_{11}=0, \Gamma_{22}=2J$. }
\label{nonr}
\end{figure}
It immediately follows that, the {excited probability} decays with time due to the dissipation of the two atoms. However, the decay behavior is also dependent on the competition between the dissipation and coupling.
For the situation of {$|\Delta|<4J$}, $P_1(t)$ will undergo small oscillation during the decay, while the oscillation will disappear as {$|\Delta|>4J$}, the different behaviors can be observed clearly in {Fig.~\ref{nonr} (a)}, where the dynamical evolution of the population is plotted for different $\Gamma_{11}$ and $\Gamma_{22}$. The change from the oscillation decay to non-oscillation decay can be physically explained by the $\mathcal{PT}$-like symmetry phase transition. To understand it intuitively, we describe the system by the non-Hermitian Hamiltonian phenomenologically by neglecting the jumping term in the master equation [Eq.~(\ref{master2})]. In this sense, the effective Hamiltonian which governs the dynamics can be written as

{\begin{equation}
H_{\rm eff}=J(\sigma_{1}^{+}\sigma_{2}^{-}+\sigma_{2}^{+}\sigma_{1}^{-})-\frac{i\Gamma_{11}}{2}|e\rangle_1\langle e|-\frac{i\Gamma_{22}}{2}|e\rangle_2\langle e|.
\end{equation}}
In the single excitation subspace, the eigen frequencies are
\begin{equation}
\omega_{\pm}=-i\frac{\delta}{4}\pm\frac{\sqrt{-K}}{4}.
\end{equation}
It shows that the system will undergo a $\mathcal{PT}$-like phase transition as $K$ moves from negative value to positive value. In the $\mathcal{PT}$-like symmetry phase with $K<0$ {($|\Delta|<4J$)}, the imaginary parts of $\omega_{\pm}$ are coincided but the real parts depart, and $M_+\sim\cos(\alpha t/2),M_-\sim\sin(\alpha t/2)$ with $\alpha=\sqrt{|K|}$. As a result, the dynamics shows an oscillation.  In the $\mathcal{PT}$-like symmetry broken phase with $K>0$ {($|\Delta|>4J$)}, the real parts of $\omega_{\pm}$ {disappear} and the imaginary parts depart with $M_+\sim\cosh(\alpha t/2),M_-\sim\sinh(\alpha t/2)$, corresponding to a pure decay dynamical behavior.

We would like to point that, in our scheme, the atom is resonant with the resonators in the waveguide, {therefore the size of each giant atom and the relative location
between them make it possible to control the effective interaction and collective/individual
dissipation. In the experiments based on superconducting qubit, the nonlocal coupling between
giant atom and the transmission line can be realized by the capacitance, and two or even more coupling points have been achieved so far~\cite{BK2019,AM2021}. On the other hand, the previous calculations show that} $\Gamma_{11}$ and $\Gamma_{22}$ can only be taken as $0$ or {$2J$}, so that the nonzero $\Delta$ only yields {$|\Delta|=2J<4J$}. That is, we can only work in the $\mathcal{PT}$-like symmetry phase, and observe the oscillation decay in our system. Meanwhile, we have used the non-Hermition Hamiltonian $H_{\rm eff}$ to analyze $\mathcal{PT}$-like phase transition physics from the viewpoint of the eigen frequency. As for the dynamics, the non-Hermition Hamiltonian only works well in the single excitation subspace, which implies that the jump term in the master equation will not take effect. But this is not the case for multiple excitations. For example, when the initial state is prepared as $|\psi(0)\rangle=|e;e\rangle$, the correct dynamics behavior can be given by the master equation but not $H_{\rm eff}$, because the trace preservation is broken by the non-Hermitian terms.

 We now consider that the second atom is excited initially, with
$|\psi_2(0)\rangle=|g;e\rangle$. Then, {the excitation probability for the same atom} becomes

{\begin{eqnarray}
P_2(t)&=&{\rm Tr}[|e\rangle_2\langle e|\rho_2(t)]\nonumber \\
&=&\frac{e^{-\frac{\delta t}{2}}}{2K^2}
[(\Delta^2-8J^2)M_+(t)-\sqrt{K}\Delta M_-(t)-16J^2],\nonumber \\
\label{trap2}
\end{eqnarray}}
where $\rho_2(t)$ is the density matrix.

We plot the comparison between $P_1(t)$ and $P_2(t)$ in {Fig.~\ref{nonr}(b)} for the achieved parameters in our giant atom scheme. It clearly shows a non-reciprocal {evolution} in that $P_1(t)\neq P_2(t)$. This fact can also be observed by the different signs before the second term of the second lines in Eqs.~(\ref{trap1}) and (\ref{trap2}). It implies that the different decay rates of the two atoms lead to the non-reciprocal excitation {evolution}.

Interestingly, dramatically different from the unidirectional transmission in most of $\mathcal{PT}$ symmetry system~\cite{PT1,PT2}, we here observe a reciprocal excitation transmission, that is,
{\begin{eqnarray}
T&=&{\rm Tr}[|e\rangle_2\langle e|\rho_1(t)]={\rm Tr}[|e\rangle_1\langle e|\rho_2(t)]\nonumber\\
&=&\frac{4J^2}{K}e^{-\frac{\delta t}{2}}(e^{\frac{Kt}{2}}+e^{-\frac{Kt}{2}}-2).
\end{eqnarray}}
{It is obvious that the excitation transmission rate $T$ depends on $K$ via $\Delta^2$, but the {excitation population} $P_1(t)$ and $P_2(t)$ depend not only on $\Delta^2$, but also $\Delta$. As a result, the latter one shows non-reciprocal behavior but the former is reciprocal.}

\section{Conclusion and Remark}
\label{con}
In conclusion, we have investigated the controllable dissipation of giant atom and its potential applications in a waveguide QED system. We show that the decay rate of the giant atom can be well controlled by changing its size, that is, the distance between two connecting points with the waveguide. With the state-to-art experimental feasibility, we propose a robust entangled state preparation scheme. {More interestingly, for two giant-atom setup, we find the non-reciprocal excitation evolution for the fixed atom  but reciprocal transmission between them.}

{We need to point out that, the controllable dissipation and effective interaction via tuning the size of the giant atom are also studied in the waveguide with linear dispersion relation~\cite{AF2018,HZ2013}. On the contrary,  we here focus on a lattice waveguide model with nonlinear dispersion relation~\cite{zhou20081}, the inter-site coupling $\xi$, which modulates the group velocity of the photons in the waveguide ($v_g=\partial \omega_k/\partial k=2\xi\sin k$), is also a sensitive parameter to control the dissipation and interaction [see Eqs.~(\ref{CA}) and~(\ref{CAS})]. Meanwhile, for our model, the waveguide forms an energy band, and the non-Markovian effect will dominant when the atom is resonant with the edge of waveguide band in energy~\cite{IV2005,IV2008}, which is not possible in the linear waveguide. The other origin of non-Markovian effect comes from the time delay of the photon transmission~\cite{LG2017, GA2019} between the two connecting points, which may appear in the waveguide with both of linear and nonlinear dispersion relation. However, these non-Markovian effects will be deeply studied in our future work.}

In addition, the realization of giant atom is not only limited in photonic waveguide, but also in acoustic system~\cite{SDbook,DM,MV2014,RM2017} and has been theoretically proposed in cold atom system~\cite{SG2019}. Therefore, we hope that our investigation has potential application in quantum information processing in these physical systems.

\begin{acknowledgments}
This work is supported by  National Natural Science Foundation of China (Grant No.~11875011,
No.~12047566 and No.~12074061).
\end{acknowledgments}

\appendix%\appendixpage
\addcontentsline{toc}{section}{Appendices}\markboth{APPENDICES}{}
\begin{subappendices}
\begin{widetext}
\section{The Master equation}
\label{master}
{In the main text, we have given the final master equation for a single giant atom setup as Eq.~(\ref{masters}), in this appendix, we will give some detailed derivations. Under the Markov approximation and working in the interaction picture, the formal master equation for a quantum open system reads~\cite{book}
\begin{equation}
\dot{\rho}(t)=-\int_{0}^{\infty}d\tau {\rm Tr}_{c}\{[H_I(t),[H_{I}(t-\tau),\rho_{c}\otimes\rho(t)]]\}.
\end{equation}}

{In our system, the interaction Hamiltonian is given in Eq.~(\ref{interaction}) {of the main text}, and the master equation yields
{\begin{eqnarray}
\dot{\rho}(t)&=&-\int_{0}^{\infty}d\tau{\rm Tr}_{c}\{[H_{I}(t),[H_{I}(t-\tau),\rho_{c}\otimes\rho(t)]]\}\nonumber \\
	&=&	-\int_{0}^{\infty}d\tau{\rm Tr}_{c}[H_{I}(t)H_{I}(t-\tau)\rho_{c}\otimes\rho(t)]
+\int_{0}^{\infty}d\tau{\rm Tr}_{c}[H_{I}(t)\rho_{c}\otimes\rho(t)H_{I}(t-\tau)]\nonumber \\
	&&+\int_{0}^{\infty}d\tau{\rm Tr}_{c}[H_{I}(t-\tau)\rho_{c}\otimes\rho(t)H_{I}(t)]-\int_{0}^{\infty}d\tau{\rm Tr}_{c}[\rho_{c}\otimes\rho(t)H_{I}(t-\tau)H_{I}(t)].
\end{eqnarray}}	
Since we are working at the zero temperature, the CRW is in the vacuum state initially, therefore, we will have ${\rm Tr}_{c}[E^{\dagger}(n_{i},t)E(n_{j},t-\tau)\rho_{c}]=0$, and the above equation becomes (going back to the Schr\"{o}dinger picture)
\begin{equation}
\dot{\rho}=-i\Omega[|e\rangle\langle e|,\rho]+(A+A^{*})\sigma^{-}\rho\sigma^{+}
-A\sigma^{+}\sigma^{-}\rho-A^{*}\rho\sigma^{+}\sigma^{-}
\end{equation}
where~\cite{GC2016}
\begin{eqnarray}
A&=&g^2\int_{0}^{\infty}d\tau e^{i\Omega\tau}{\rm Tr}_{c}[\sum_{i,j}E(n_{i},t)E^{+}(n_{j},t-\tau)\rho_{c}]\nonumber\\
	&=&	g^2\sum_{i,j}\int_{0}^{\infty}d\tau e^{i\Omega\tau}{\rm Tr}[E(n_{i},t)E^{+}(n_{j},t-\tau)\rho_{c}]\nonumber \\
	&=&	g^2\sum_{i,j}\int_{0}^{\infty}d\tau \frac{e^{i\Omega\tau}}{N_{c}}\times{\rm Tr}[\sum_{k,k'}e^{-i\omega_{k}t}e^{ikn_{i}}a_{k}e^{i\omega_{k'}(t-\tau)}e^{-ik'n_{j}}
a_{k'}^{\dagger}\rho_{c}]\nonumber \\
	&=&	g^2\sum_{i,j}\int_{0}^{\infty}d\tau\frac{1}{N_{c}}\sum_{k}[e^{-i(\omega_{k}-\Omega)\tau}
e^{-ik(n_{j}-n_{i})}]\nonumber \\
	&=&	g^2\sum_{i,j}\int_{0}^{\infty}d\tau\frac{1}{N_{c}}\sum_{n=0}^{N_{c}-1}e^{-i\Delta_c\tau}e^{\frac{-2\pi i(n_{j}-n_{i})n}{N_{c}}}e^{2i\xi\cos(\frac{2\pi n}{N_{c}})\tau}\nonumber\\
	&=&	g^2\sum_{i,j}\int_{0}^{\infty}d\tau\frac{e^{-i\Delta_c\tau}}{N_{c}}\sum_{n=0}^{N_{c}-1}e^{\frac{-2\pi i(n_{j}-n_{i})n}{N_{c}}}\times\sum_{m=-\infty}^{\infty}i^{m}J_{m}(2\xi\tau)e^{i2\pi nm/N_{c}}\nonumber \\
	&=&	g^2\sum_{i,j}\int_{0}^{\infty}d\tau e^{-i\Delta_c\tau}i^{|n_{i}-n_{j}|}J_{|n_{i}-n_{j}|}(2\xi\tau)\nonumber \\
	&=&	g^2\sum_{i,j}\frac{1}{2\xi}e^{\frac{i\pi|n_{i}-n_{j}|}{2}}\nonumber \\&=&\frac{g^2}{\xi}(1+e^{i\pi|n_{1}-n_{2}|/2}).
\end{eqnarray}
In the above calculations, we have considered that the giant atom is resonant with the bare cavity ($\Delta_c:=\omega_c-\Omega=0$), and used the formular
\begin{equation}
\int_{0}^{\infty}d\tau J_{m}(a\tau)=\frac{1}{|a|}.
\end{equation}}
\section{Numerical simulation for single giant atom system}
\label{numer}
{In this appendix, we will outline the procedure of the numerical simulation for the dissipation of single giant atom. In the main text, we have analytically derived the master equation under the Markov approximation by considering the waveguide as a structured environment. The underlying physics behind Markov approximation is that the environment {loses} its memory and {keeps} in its initial state during the time evolution. Physically speaking, {an} excited giant atom will decay to the ground state along with the emission of photon. The emitted photon will then {travel} along the waveguide. To guarantee the validity of the Markov approximation, the emitted photon must leave the atomic regime and never come back. To fulfill such condition, we choose the atomic frequency to be resonant with the bare resonator, such that the group velocity of the emitted photon achieves its maximum value, which {makes} the photon leave the atom-waveguide connecting points fast. Meanwhile, we try to enlarge the length of the waveguide and induce a small decay for each resonator to stop the emitted photon {from} going back the atomic regime. In the numerical simulation, we adopt a non-Hermitian manner by phenomenologically {expressing} the Hamiltonian as
\begin{eqnarray}
H=(\omega_c-i\kappa)\sum_j
a_j^{\dagger}a_j-\xi\sum_j
(a_j^\dagger a_{j+1}+{\rm H.c.})+\Omega|e\rangle\langle e|+g[(a_{n_1}^\dagger+a_{n_2}^\dagger)\sigma^{-}+{\rm H.c.}],
\end{eqnarray}
where $\kappa$ is the decay rate for each resonator in the waveguide. Then, the excited amplitude of the giant atom is numerically obtained as
\begin{equation}
P(t)=|\langle\psi(0)|e^{-iHt}|\psi(0)\rangle|^2
\end{equation}
where $|\psi(0)\rangle=|e,G\rangle$ represents that the giant atom is in the excited state while all of the resonators in the waveguide are in their vacuum states. In the numerical simulation {(see Fig.~\ref{atomdecay} in the main text)}, we have chosen the length of the waveguide as $N_c=4000$, and the decay rate for each resonator as $\kappa/\xi=6\times10^{-3}$. We have also checked that, for  $N_c=3000-5000$ and $\kappa/\xi=3\times10^{-3}-1.2\times10^{-2}$, the numerical results nearly unchange.}
\end{widetext}
\end{subappendices}


\begin{thebibliography}{99}
\bibitem{TP2003}T. Petroskya and S. Subbiaha, Physical E {\bf 19}, 230 (2003).

\bibitem{AF2014}A. F. Kockum, P. Delsing and G. Johansson,
Phys. Rev. A {\bf 90}, 013837 (2014).

\bibitem{LG2017}L. Guo, A. Grimsmo, A. F. Kockum, M. Pletyukhov and
G. Johansson,  Phys. Rev. A {\bf 95}, 053821 (2017).

\bibitem{AF2018}A. F. Kockum, G. Johansson and F. Nori,
Phys. Rev. Lett. {\bf 120}, 140404  (2018).

\bibitem{PT}P. T\"{u}rschmann, H.  L. Jeannic, S. F. Simonsen, H. R. Haakha,
S. G\"{o}tzinger, V. Sandoghdar, P. Lodahl and N. Rotenberg,
 Nanophotonics {\bf 8}, 1641 (2019).

\bibitem{GA2019}G. Andersson, B. Suri, L. Guo, T. Aref and P. Delsing, Nat. Phys. {\bf 15},
1123 (2019).

\bibitem{AG2019}A. G.-Tudela, C. S. Mu\~{n}oz and J. I. Cirac,
Phys. Rev. Lett. {\bf 122}, 203603 (2019).


\bibitem{LG2019}L. Guo, A. F. Kockum, F. Marquardt and G. Johansson, Phys. Rev.
Research {\bf 2}, 043014 (2020).

\bibitem{SG2019}S. Guo, Y. Wang, T. Purdy and J. Taylor, Phys. Rev. A {\bf 102}, 033706 (2020).

\bibitem{BK2019}B. Kannan, {\it et al}., Nature {\bf 583}, 775-779 (2020).

\bibitem{AM2021}{A. M. Vadiraj, A. Ask, T. G. McConkey, I. Nsanzineza,
C. W. S. Chang, A. F. Kockum and C. M. Wilson, Phys. Rev. A {\bf 103}, 023710 (2021).}

\bibitem{HJ2008}H. J. Kimble, Nature {\bf 453}, 1023 (2008).

\bibitem{Fan2005} J.-T. Shen and S. Fan, Phys. Rev. Lett. {\bf 95}, 213001 (2005).

\bibitem{DE2007}D. E. Chang, A. S. S{\o}rensen, E. A. Demler and M. D. Lukin, Nat. Phys. {\bf 3}, 807 (2007).

\bibitem{PL2010}P. Longo, P. Schmitteckert and K. Busch, Phys. Rev. Lett. {\bf 104}, 023602 (2010).

\bibitem{TS2018}T. Shi, Y. Chang and J. J. Garc\'{i}a-Ripoll, Phys. Rev. Lett. {\bf 120}, 153602 (2018).

\bibitem{Zhou2008} L. Zhou, Z. R. Gong, Y. X. Liu, C. P. Sun and F. Nori,
Phys. Rev. Lett. {\bf 101}, 100501 (2008).

\bibitem{Zhou2013} L. Zhou, L. P. Yang, Y. Li and C. P. Sun, Phys. Rev. Lett. {\bf 111}, 103604 (2013).

\bibitem{Wang2014} Z. H. Wang, L. Zhou, Y. Li and C. P. Sun, Phys. Rev.  A {\bf 89}, 053813 (2014).

\bibitem{GC2016}G. Calaj\'{o}, F.  Ciccarello, D. Chang and P. Rabl,
Phys. Rev. A {\bf 93}, 033833 (2016).

\bibitem{HZ2013}H. Zheng and H. U. Baranger, Phys. Rev. Lett. {\bf 110}, 113601 (2013).

\bibitem{ES2013}E. Shahmoon and G. Kurizki, Phys. Rev. A {\bf 87}, 033831 (2013).

\bibitem{PF2016}P. Facchi, M. S. Kim, S. Pascazio, F. V. Pepe, D. Pomarico and T. Tufarelli, Phys. Rev. A {\bf 94}, 043839 (2016).

\bibitem{Zhao2020}{W. Zhao and Z. Wang, Phys. Rev. A {\bf 101}, 053855 (2020).}

\bibitem{atomic1} D. W. Wang, H. T. Zhou, M. J. Guo, J. X. Zhang, J. Evers and S. Y. Zhu, Phys. Rev. Lett. {\bf 110}, 093901 (2013).

\bibitem{atomic2}J.-H. Wu, M. Artoni and G. C. La Rocca,  Phys. Rev. Lett. {\bf 113}, 123004 (2014).

\bibitem{PT1}B. Peng, S. K. \"{o}zdemir, F. Lei, F. Monifi, M. Gianfreda,
G. L. Long, S. Fan, F. Nori, C. M. Bender and L. Yang, Nat. Phys. {\bf 10}, 394 (2014).

\bibitem{PT2} L. Chang, X. Jiang, S. Hua, C. Yang, J. Wen, L.
Jiang, G. Li, G. Wang and M. Xiao, Nat. Photonics {\bf 8}, 524 (2014).

\bibitem{book}H. Breuer and F. Petruccione, \textit{The Theory of Open Quantum Systems}, (Oxford: Oxford University Press 2002)

\bibitem{SH2015}{S. Hacohen-Gourgy, V. V. Ramasesh, C. D. Grandi, I.
Siddiqi and S. M. Girvin, Phys. Rev. Lett. {\bf 115}, 240501 (2015).}

\bibitem{PF2017}{P. Forn-D\'{\i}az, J. J. Garc\'{\i}a-Ripoll, B. Peropadre, J.-L. Orgiazzi,
M. A. Yurtalan, R. Belyansky, C. M. Wilson and A. Lupascu,
Nat. Phys. {\bf 13}, 39 (2017).}

\bibitem{FY2017}{F. Yoshihara, T. Fuse, S. Ashhab, K. Kakuyanagi, S. Saito and
K. Semba, Nat. Phys. {\bf 13}, 44 (2017).}

\bibitem{AJ2017}{A. J. Keller, P. B. Dieterle, M. Fang, B. Berger,  J. M. Fink and O. Painter, Appl. Phys. Lett. {\bf 111}, 042603 (2017).}

\bibitem{JJ2019}{J. J. Burnett, A. Bengtsson, M. Scigliuzzo, D. Niepce, M. Kudra, P. Delsing and
J. Bylander, npj Quantum Information {\bf 5}, 54 (2019). }

\bibitem{IV2005}I. de Vega, D. Alonso and P. Gaspard, Phys. Rev. A {\bf 71}, 023812 (2005).

\bibitem{IV2008}I. de Vega and D. Alonso, Phys. Rev. A {\bf 77}, 043836 (2008).

\bibitem{zhou20081}{L. Zhou, H. Dong, Y.-x. Liu, C. P. Sun and F. Nori, Phys. Rev. A {\bf 78}, 063827 (2008).}

\bibitem{SDbook}S. Datta, {\it Surface Acoustic Wave Devices}, (Prentice-Hall, Englewood Cliffs, NJ, 1986).

\bibitem{DM}D. Morgan, {\it Surface Acoustic Wave Filters}, 2nd ed. (Academic,
Amsterdam, 2007).

\bibitem{MV2014}M.  V. Gustafsson, T. Aref,  A. F. Kockum, M. K. Ekstr\"{o}m,
G. Johansson and P. Delsing, Science {\bf 346}, 207 (2014).

\bibitem{RM2017}R. Manenti, A. F. Kockum, A. Patterson, T. Behrle, J. Rahamim,
G. Tancredi, F. Nori and P. J. Leek, Nat. Comm. {\bf 8}, 975 (2017).



\end{thebibliography}
\end{document}